\documentclass[aps,prl,preprint,groupedaddress]{revtex4}

\usepackage{amssymb}
\usepackage{epstopdf}
\usepackage{graphicx}
\usepackage{bm}
\newcommand{\bra}[1]{\langle #1|}
\newcommand{\ket}[1]{|#1\rangle}

\begin{document}


\title{Quantum interference and Bayes theorem}


\author{Filippo Neri}
\thanks{Research supported in part by the US Department of Energy under contract W-7405-Eng-36}
\affiliation{Los Alamos National Laboratory, Los Alamos, NM 87545, USA.}


\date{\today}

\begin{abstract}
I give a simple discussion of the accuracy limits of quantum interferometry using
an elementary application of Bayes theorem. This work is motivated, in part, by the
appearance in the literature of partially incorrect results. My aim is to
give consistent results and a compact, if formal, discussion of the ideas
of Bayesian analysis of quantum phase measurements.
\end{abstract}


\maketitle


Using the methods of Refs.~\cite{Hradil-2003-, Hradil95} and an elementary
application of Bayes theorem, we can
derive the probability of the phase having the value $\theta$, if the input state
 $\ket{\Psi}$ is observed at the output of a Mach-Zehnder interferometer:
\begin{equation}
P(\theta) = | \bra{\Psi} e^{i \theta \hat{L}_2} \ket{\Psi} |^{2} = 
1 - {\theta}^2 \left(\bra{\Psi}{{\hat{L}_2}^2}\ket{\Psi} 
-\bra{\Psi}{{\hat{L}_2}}\ket{\Psi}^2 \right) + ... ,
\end{equation}
where
\begin{equation}
\hat{L}_2 = {{1}\over{2 i}} (\hat{a}^{\dagger} \hat{b} - \hat{b}^{\dagger} \hat{a}).
\end{equation}
We have used an expansion around zero, because we assume that prior
knowledge restricts the phase to the neighborhood of zero. 
We will also assume that the observation is further
repeated, narrowing the uncertainty in $\theta$.

If $M$ copies (we assume  that $M$ is large) of the state are sent through
the interferometer, with a null
result every time, the probability of the phase being $\theta$ is given by
the expression
\begin{equation}
P(\theta)^M \approx
e^{-M {\theta}^2
\left(\bra{\Psi}{{\hat{L}_2}^2}\ket{\Psi} 
-\bra{\Psi}{{\hat{L}_2}}\ket{\Psi}^2 \right)
}
= e^{-{1\over 2} {{\theta^2} / {{\Delta\theta}^2}}},
\end{equation}
where 
\begin{equation}
 {{1} \over {{\Delta\theta}^2}}
 = {{M}\over{2}} \left( {\bra{\Psi}{4 {\hat{L}_2}^2}\ket{\Psi}} -
 4 \bra{\Psi}{{\hat{L}_2}}\ket{\Psi}^2 \right).
 \label{DTH2}
\end{equation}
We can write the operator $4 {\hat{L}_2}^2$ as
\begin{equation}
4 {\hat{L}_2}^2 =
-{\hat{b}}^{{\dagger 2}} {\hat{a}}^2
-{\hat{a}}^{{\dagger 2}} {\hat{b}}^2
+2 {\hat{a}}^{{\dagger}}{\hat{b}}^{{\dagger}}{\hat{b}}{\hat{a}} 
+ {\hat{a}}^{{\dagger}}{\hat{a}}+{\hat{b}}^{{\dagger}}{\hat{b}}.
\label{4F22}
\end{equation}
The form of Eq.~(\ref{4F22}) is useful, because the operators are normal-ordered.
Assuming that $\ket{\Psi}$ is the direct product of two Fock states
\begin{equation}
\ket{\Psi}  = \ket{n_a} \otimes \ket{n_b},
\end{equation}
using Eq.~(\ref{DTH2}) and (\ref{4F22}), we obtain the result
\begin{equation}
 {{1} \over {{\Delta\theta}^2}}
 = {{M}\over{2}}(2n_a n_b + n_a + n_b).
\end{equation}
If we define as $N$ the total number of photons, then $M=N/(n_a+n_b)$ and
\begin{equation}
 {{1} \over {{\Delta\theta}^2}}
 = {{N}\over{2n_a+2n_b}}(2n_a n_b + n_a + n_b).
 \label{thNN}
\end{equation}
Special cases of Eq.~(\ref{thNN}) are the following:
If $n_a=1$ and $n_b=0$,
\begin{equation}
 {{1} \over {{\Delta\theta}^2}}
 = {{N}\over{2}}.
 \label{SQL}
\end{equation}
If  $n_a=n_b=1$,
\begin{equation}
 {{1} \over {{\Delta\theta}^2}}
 = {{N}}.
 \label{SQL2}
\end{equation}
If  $n_a=n_b=n$,
\begin{equation}
 {{1} \over {{\Delta\theta}^2}}
 = {{N}\over{2}}(n+1).
 \label{SQL3}
\end{equation}
Eq~(\ref{SQL}) is (our version of) the Standard Quantum Limit (SQL). Eq~(\ref{SQL2})
shows that correlated photons pairs beat the SQL by 3 dB. (Ref.~\cite{Hradil-2003-}
incorrectly states that photons pairs are not better than SQL. This ``result'' is obtained by
using a large-$l$ asymptotic expansion of $P_l(\cos\theta)$ for $l=1$. This is equivalent to
dropping the $1$ in Eq.~(\ref{SQL3}), because is negligible compared to $n$, and then
setting $n$ to 1. My aim in this note is not to pick nits in Ref.~\cite{Hradil-2003-},
but to show how its results can be extended to states different from Fock states.
However, I have to point out that Eq.~(39) in Ref.~\cite{Hradil-2003-} is also incorrect:
the asymptotic expansion used is only valid for large $l$ {\it and} small argument.)

I  will discuss next states of the form
\begin{equation}
\ket{\Psi}  = \ket{\psi_a} \otimes \ket{b e^{i\phi}},
\label{PSI}
\end{equation}
where $\ket{b e^{i\phi}}$ is a coherent state:
\begin{equation}
\hat{b}\ket{b e^{i\phi}} = b e^{i\phi} \ket{b e^{i\phi}},
\end{equation}
and $\ket{\psi_a}$ is an arbitrary state of the $\hat{a}$ photons.
Using the state of Eq.~(\ref{PSI}) in Eq.~(\ref{DTH2}), we get
\begin{equation}
 {{1} \over {{\Delta\theta}^2}}
 = {{M}\over{2}} \left[ \bra{\psi_a}
 {\left( -b^2 e^{-i2\phi} \hat{a}^2 -b^2 e^{i2\phi} \hat{a}^{\dagger 2} 
 +2 b^2 \hat{a}^{\dagger}\hat{a} + \hat{a}^{\dagger}\hat{a} + b^2
\right)} \ket{\psi_a}
+b^2 \bra{\psi_a} \left(e^{-i \phi} \hat{a} - e^{i \phi} \hat{a}^{\dagger} 
 \right) \ket{\psi_a}^2 \right].
\end{equation}
Since the average numbers of photons in the state $\ket{\psi_a}$ is
\begin{equation}
n_a = \bra{\psi_a} \hat{a}^{\dagger}\hat{a} \ket{\psi_a},
\end{equation}
and the average number in the state $\ket{b e^{i\phi}}$ is
\begin{equation}
n_b = b^2,
\end{equation}
the (average) total number
of photons is
\begin{equation}
N = M \left( n_a + n_b \right).
\end{equation}
We can then write the general result in the form
\begin{equation}
 {{1} \over {{\Delta\theta}^2}}=
 {{N \left[b^2 \bra{\psi_a} \left(e^{-i \phi} \hat{a} - e^{i \phi} \hat{a}^{\dagger} 
 \right) \ket{\psi_a}^2
 -b^2 \bra{\psi_a} \left(e^{-i \phi} \hat{a} - e^{i \phi} \hat{a}^{\dagger} 
 \right)^2 \ket{\psi_a} + n_a \right]
 }\over{2 \left( n_a + n_b \right)}}.
 \label{1TH2COH}
\end{equation}
We define, for an arbitrary $\beta$,
\begin{equation}
\hat{Q}(\beta) = e^{-i \beta} \hat{a} + e^{i \beta} \hat{a}^{\dagger}.
\label{QU}
\end{equation}
Eq.~(\ref{1TH2COH}) can then be rewritten as
\begin{equation}
 {{1} \over {{\Delta\theta}^2}}=
 {{N \left\{n_b \bra{\psi_a}\left[\Delta\hat{Q}(\phi-{{\pi}\over{2}}) \right]^2
 \ket{\psi_a} + n_a \right\} }\over{2 \left( n_a + n_b  \right) }},
\label{1TH2Q}
\end{equation}
where
\begin{equation}
\Delta\hat{Q}(\beta)=\hat{Q}(\beta)-
\bra{\psi_a}\hat{Q}(\beta)\ket{\psi_a}.
\label{DQU}
\end{equation}

Eq.~(\ref{1TH2Q}) is the main result of this note.
It is valid for an arbitrary state at the $a$ input---if a coherent state
is present at the $b$ input. It is expressed in term only of the average
photon numbers and of the
standard deviation of the quadrature amplitude
$\hat{Q}(\phi-{{\pi}\over{2}})$ in the state $\ket{\psi_a}$. This
result is exact---in the sense that we have not made any
approximation in Eq.~(\ref{DTH2}).
In particular, we have not made the assumption that $b^2$ is
large compared to the photon number in the state $\ket{\psi_a}$:
the expression is not linearized.

If the state $\ket{\psi_a}$ is also a
coherent state, Eq.~(\ref{1TH2Q}) reduces exactly to Eq~(\ref{SQL}): the SQL. In
particular this is true if $\ket{\psi_a}$ is the vacuum state (the vacuum state
{\it is} a coherent state!)

If the state $\ket{\psi_a}$ is a squeezed vacuum state, 
\begin{equation}
\ket{\psi_a} =
e^{(z^* \hat{a}^2 - z \hat{a}^{\dagger 2})/2} \ket{0_a},
\end{equation}
where
\begin{equation}
z=r e^{i \psi},
\end{equation}
we have
\begin{equation}
 {{1} \over {{\Delta\theta}^2}}=
 {{N \left\{n_b \left[\cosh 2r -\sinh 2r \cos(\psi - 2\phi +\pi)
 \right] + \sinh^2 r \right\} }\over{2 \left(n_b  + \sinh^2 r \right) }}.
\label{1TH2S}
\end{equation}
(For a discussion of squeezed states and a derivation of the result
used above for the standard deviation of the quadrature amplitude,
see Chapter 21 of Ref.~\cite{Man95}.)
Eq.~(\ref{1TH2S}) has a maximum at $\psi=2\phi$ (we assume $r > 0$),
where it has the form
\begin{equation}
{{1} \over {{\Delta\theta}^2}}=
 {{N \left( n_b e^{2 r}
+ \sinh^2 r \right) }\over{2 \left(n_b  + \sinh^2 r \right) }}.
\label{1TH2M}
\end{equation}
If we now take $n_b \gg \sinh^2 r$, we see that we can beat the
SQL by a factor $e^{2 r}$.
This is, of course, a well-known result (originally derived in Ref.~\cite{Caves81}). The point
of this note was to derive it using the methods of Ref.~\cite{Hradil-2003-}.

\bibliography{QBint}

\end{document}